\documentclass[preprint2]{aastex}
\shorttitle{NGC 2024 FIR 5/6 Radio Imaging}
\shortauthors{Choi, Lee, and Kang}

\usepackage{times}
\slugcomment{To appear in the Astrophysical Journal}

\begin{document}

\fontsize{10}{10.6}\selectfont

\title{Radio Imaging of the NGC 2024 FIR 5/6 Region:
       a Hypercompact H {\small II} Region Candidate in Orion} 
\author{\sc Minho Choi$^{1,2}$, Jeong-Eun Lee$^3$, and Miju Kang$^1$} 
\affil{$^1$ Korea Astronomy and Space Science Institute,
            776 Daedeokdaero, Yuseong, Daejeon 305-348, Republic of Korea;
            minho@kasi.re.kr \\
       $^2$ Astronomy and Space Science Major,
            University of Science and Technology,
            217 Gajeongro, Yuseong, Daejeon 305-350, Republic of Korea \\
       $^3$ Department of Astronomy and Space Science,
            Kyung Hee University, Yongin, Gyeonggi 446-701, Republic of Korea}
\setcounter{footnote}{3}

\begin{abstract}
\fontsize{10}{10.6}\selectfont
The NGC 2024 FIR 5/6 region was observed in the 6.9 mm continuum
with an angular resolution of about 1.5 arcsec.
The 6.9 mm continuum map shows four compact sources, FIR 5w, 5e, 6c, and 6n,
as well as an extended structure of the ionization front
associated with the optical nebulosity.
FIR 6c has a source size of about 0.4 arcsec or 150 AU.
The spectral energy distribution (SED) of FIR 6c is peculiar:
rising steeply around 6.9 mm and flat around 1 mm.
The possibility of a hypercompact H {\small II} region is explored.
If the millimeter flux of FIR 6c comes from hot ionized gas
heated by a single object at the center,
the central object may be a B1 star
of about 5800 solar luminosities and about 13 solar masses.
The 6.9 mm continuum of FIR 6n may be a mixture
of free--free emission and dust continuum emission.
Archival data show that both FIR 6n and 6c exhibit water maser activity,
suggesting the existence of shocked gas around them.
The 6.9 mm continuum emission from FIR 5w
has a size of about 1.8 arcsec or 760 AU.
The SEDs suggest
that the 6.9 mm emission of FIR 5w and 5e comes from dust,
and the masses of the dense molecular gas
are about 0.6 and 0.5 solar masses, respectively.
\end{abstract}


\section{INTRODUCTION}

The NGC 2024 region in the Orion B giant molecular cloud
is an active site of star formation
at a distance of 415 pc from the Sun
(Anthony-Twarog 1982; Comer{\'o}n et al. 1996;
Skinner et al. 2003; Rodr{\'\i}guez et  al. 2003).
There is a molecular ridge running in the north--south direction
that contains the dense cores FIR 1--7
(Thronson et al. 1984; Barnes et al. 1989;
Mezger et al. 1988, 1992; Visser et al. 1998).
Some of the dense cores contain protostars
and exhibit star formation activities
such as collimated outflows and an H$_2$O maser
(Genzel \& Downes 1977; Richer et al. 1989, 1992;
Richer 1990; Chernin 1996; Chandler \& Carlstrom 1996).

The molecular ridge is interacting with an expanding H {\small II} region
(Gaume et al. 1992).
The bright emission from the H {\small II} region makes it difficult
to study the young stellar objects (YSOs) in the radio and IR bands.
Interferometric observations can filter out
the extended emission from the H {\small II} region
and provide detailed images of the compact objects in this region
(Wilson et al. 1995; Chandler \& Carlstrom 1996).

The dense core NGC 2024 FIR 6 may contain a protostar
and was resolved into two objects in millimeter images
(Chandler \& Carlstrom 1996; Visser et al. 1998; Lai et al. 2002).
The nature of the brighter source, FIR 6c, is not clear.
The other member of the binary system, FIR 6n,
is the driving source of a bipolar outflow
with a very short (400 yr) timescale (Richer 1990; Alves et al. 2011).

The dense core FIR 5 may contain embedded protostars
though the evolutionary status is not clear
(Chandler \& Carlstrom 1996; Visser et al. 1998).
The interferometric imaging in the 3.1 mm continuum
by Wiesemeyer et al. (1997)
showed that FIR 5 is a binary system.
More sensitive imaging in the 1.3 mm continuum by Lai et al. (2002) showed
that the FIR 5 system is a complex of several compact sources.
The brightest one, FIR 5w, is probably the driving source
of the highly collimated unipolar outflow flowing to the south
(Alves et al. 2011).

In this paper, we present the results of
our observations of the NGC 2024 FIR 5/6 region
in the 6.9 mm continuum with the Very Large Array (VLA)
of the National Radio Astronomy Observatory (NRAO).
We describe our radio continuum observations in Section 2.
Two archival data sets with the H$_2$O maser observations
are described in Section 3.
In Section 4, we report the results of the radio imaging.
In Section 5, we discuss the star-forming activities in the FIR 5/6 region.
A summary is given in Section 6.

\section{OBSERVATIONS}

The NGC 2024 FIR 5/6 region was observed using the VLA
in the standard $Q$-band continuum mode (43.3 GHz or $\lambda$ = 6.9 mm).
Twenty-five antennas were used
in the D-array configuration on 2003 March 30.
The phase-tracking center was
$\alpha_{2000}$ = 05$^{\rm h}$41$^{\rm m}$44\fs69,
$\delta_{2000}$ = --01\arcdeg55$'$52\farcs5,
which is the midpoint between FIR 5 and 6.

The phase was determined
by observing the nearby quasar 0532+075 (PMN J0532+0732).
The flux was calibrated
by setting the flux density of the quasar 0542+498 (3C 147) to 0.72 Jy,
which is the flux density measured within a day of our observations
(VLA Calibrator Flux Density Database%
\footnote{See http://aips2.nrao.edu/vla/calflux.html.}).
Comparison of the amplitude gave a flux density of 1.04 Jy for 0532+075.
To avoid the degradation of sensitivity owing to pointing errors,
pointing was referenced
by observing the calibrators in the $X$ band ($\lambda$ = 3.6 cm).
This referenced pointing was performed
about once an hour and just before observing the flux calibrator.

Maps were made using a CLEAN algorithm.
With a natural weighting,
the 6.9 mm continuum data produced a synthesized beam
of 2\farcs2 $\times$ 1\farcs6 in full width at half-maximum (FWHM).

\section{ARCHIVAL DATA}

In addition to our data, we also analyzed two VLA data sets
retrieved from the NRAO Data Archive System.
We searched for H$_2$O maser observations of the FIR 5/6 region
to investigate the source identification issue raised previously
(see Section 4.3.4 of Furuya et al. 2003).
In this paper, we present the results of the NRAO project AC 443.

\begin{figure*}[!t]
\epsscale{2}
\plotone{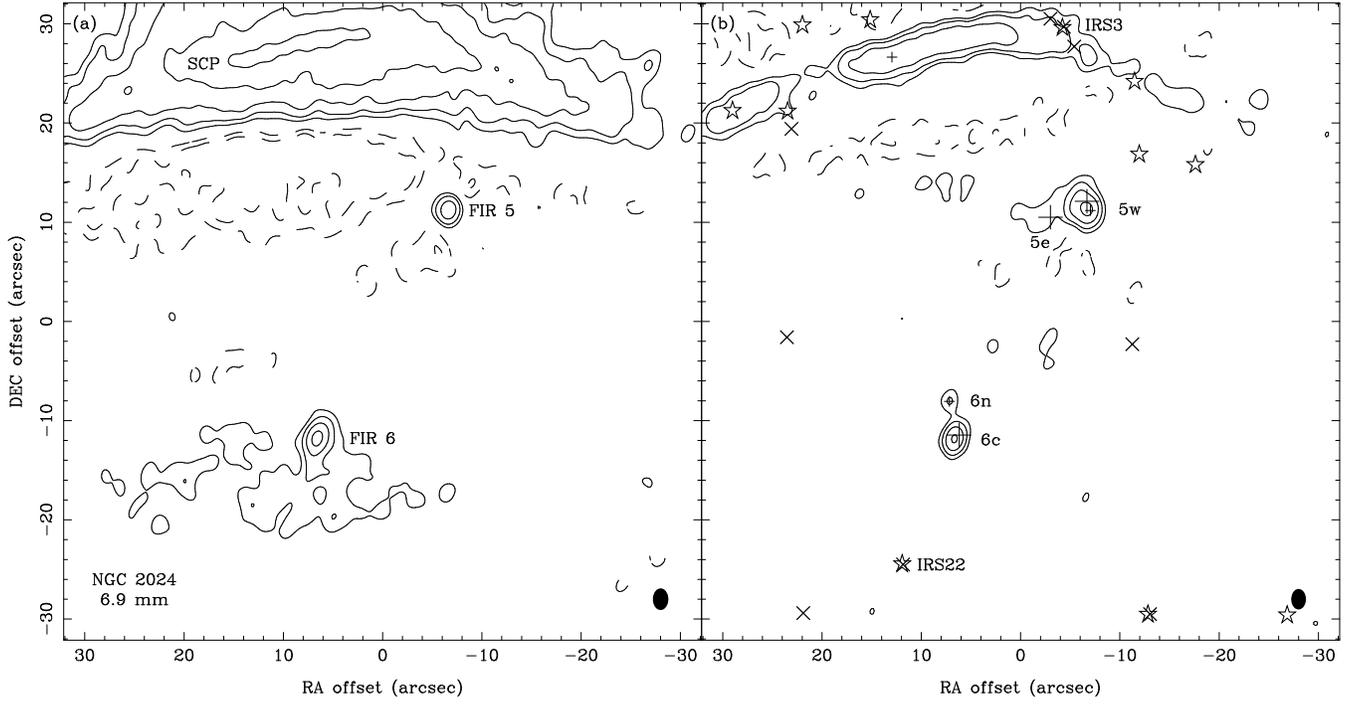}
\caption{\small\baselineskip=0.825\baselineskip
Maps of the $\lambda$ = 6.9 mm continuum emission
toward the NGC 2024 FIR 5/6 region.
The contour levels are 1, 2, 4, 8, and 16 $\times$ 0.3 mJy beam$^{-1}$.
Dashed contours are for negative levels.
(a)
Map made with the whole data set.
The rms noise is 0.11 mJy beam$^{-1}$.
Shown in the bottom right-hand corner is the synthesized beam:
FWHM = 2\farcs2 $\times$ 1\farcs6 with P.A. = 0$^\circ$.
(b)
Map made with the data having the $uv$ length larger than 12 k$\lambda$
and with a natural weighting.
The rms noise is 0.12 mJy beam$^{-1}$.
The synthesized beam has
FWHM = 2\farcs1 $\times$ 1\farcs5 with P.A. = 0$^\circ$.
Large plus signs:
the 3.1 mm continuum sources (Wiesemeyer et al. 1997).
Small plus signs:
the 3.6 cm continuum sources (Rodr{\'\i}guez et al. 2003).
Star symbols:
near-IR sources listed in the 2MASS Point Source Catalogue.
IRS 3 and IRS 22 are labeled.
Crosses:
X-ray sources (Skinner et al. 2003).}
\end{figure*}

\begin{deluxetable}{lcccrcc}
\tabletypesize{\small}
\tablecaption{NGC 2024 Continuum Source Parameters}%
\tablewidth{0pt}
\tablehead{
& \multicolumn{2}{c}{Peak Position\tablenotemark{a}}
&& \multicolumn{2}{c}{6.9 mm Flux Density\tablenotemark{b}}
& \colhead{Associated} \\
\cline{2-3} \cline{5-6}
\colhead{Source}
& \colhead{$\alpha_{\rm J2000.0}$} & \colhead{$\delta_{\rm J2000.0}$}
&& \colhead{Peak} & \colhead{Total} & \colhead{Objects\tablenotemark{c}}}%
\startdata
FIR 5w & 05 41 44.24 & --01 55 41.2 &&  3.31 $\pm$ 0.14 & 7.8 $\pm$ 0.7
       & LCGR 4, VLA 10 \\
FIR 5e & 05 41 44.56 & --01 55 42.3 &&  0.64 $\pm$ 0.13 & 1.3 $\pm$ 0.2
       & LCGR 6 \\
FIR 6c & 05 41 45.14 & --01 56 04.4 &&  2.99 $\pm$ 0.14 & 3.8 $\pm$ 0.4
       & \ldots \\
FIR 6n & 05 41 45.17 & --01 56 00.5 &&  0.71 $\pm$ 0.13 & Unresolved
       & VLA 14 \\
SCP    & 05 41 45.47 & --01 55 25.5 && 10.24 $\pm$ 0.21
       & $>$ 610\tablenotemark{d}
       & VLA 17 \\
\enddata\\
\tablecomments{For the compact sources (FIR 5w, 5e, 6c, and 6n),
               the source parameters are based on the optimized map
               (Figure 1(b)).
               For SCP, they are based on the map
               made without the $uv$ restriction (Figure 1(a)).}%
\tablenotetext{a}{Units of right ascension are hours, minutes, and seconds,
                  and units of declination are degrees, arcminutes,
                  and arcseconds.}%
\tablenotetext{b}{Flux densities are in mJy beam$^{-1}$ or mJy,
                  corrected for the primary beam response.}%
\tablenotetext{c}{Source numbers from LCGR: Lai et al. (2002);
                  VLA: Rodr{\'\i}guez et al. (2003).}%
\tablenotetext{d}{SCP extends beyond the boundary of the field of view.}%
\end{deluxetable}

Twenty-seven antennas were used
in the B-array configuration on 1995 November 17
and in the C-array configuration on 1996 March 15.
For the H$_2$O $6_{16} \rightarrow 5_{23}$ line (22.235077 GHz),
the spectral window was set to have 128 channels
with a channel width of 0.049 MHz,
giving a velocity resolution of 0.66 km s$^{-1}$.
The phase-tracking center was
$\alpha_{1950}$ = 05$^{\rm h}$39$^{\rm m}$13\fs6,
$\delta_{1950}$ = --01\arcdeg57$'$30\farcs0,
which is within 1$''$ from the 6.9 mm peak position of FIR 6c.
The phase calibrator was the quasar 0607--085 (QSO B0605--0834).
With a uniform weighting,
the B-array data produced a synthesized beam
of FWHM = 0\farcs38 $\times$ 0\farcs29
with a position angle (P.A.) of 0$^\circ$,
and the C-array data produced a synthesized beam
of FWHM = 1\farcs01 $\times$ 0\farcs80 with P.A. = --1$^\circ$.

\section{RESULTS}

\subsection{Continuum Emission}

\begin{figure*}[!t]
\epsscale{2}
\plotone{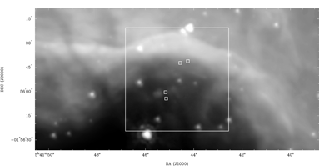}
\centerline{\scriptsize [See http://minho.kasi.re.kr/Publications.html
for the original high-quality figure.]}
\vspace{-\baselineskip}
\caption{\small\baselineskip=0.825\baselineskip
Color composite image
of the 3.6 (blue), 4.5 (green), and 5.8 (red) $\mu$m bands
toward the NGC 2024 region from the {\it Spitzer} data archive.
Small squares:
the 6.9 mm continuum sources.
Large square:
the VLA field of view at 6.9 mm (the area shown in Figure 1).}
\end{figure*}

Figure 1 shows maps of the 6.9 mm continuum
toward the NGC 2024 FIR 5/6 region,
and Table 1 lists the continuum source parameters.
Figure 1(a) shows the map made with the whole data set.
In addition to the compact sources associated with FIR 5 and 6,
the northern edge of the map is dominated by an extended source.
This structure is a part of the southern compact peak (SCP)
that has been imaged at longer wavelengths
(Crutcher et al. 1986; Barnes et al. 1989; Gaume et al. 1992).
SCP is an ionization front associated with the optical/IR nebulosity
and several IR/X-ray sources (Figures 1(b) and 2).

Despite the name, SCP is a very extended structure
with respect to our resolution
and produces spurious features of stripes oriented east--west.
In Figure 1(a), the area around FIR 5 is affected by a negative stripe,
and the FIR 6 area is affected by a positive one.
Since most of the flux of the SCP is in the extended structure,
the unwanted effects of the SCP on neighboring areas can be reduced
by excluding the visibility data of short baselines.
Imaging can be done using the visibility data
with the $uv$ length larger than a certain value ($R_{uv}^{\rm min}$).
Several values of $R_{uv}^{\rm min}$ were tried,
and the location and intensity of the spurious stripe features were examined.
We found that an optimal value for the imaging of FIR 5/6
is $R_{uv}^{\rm min}$ = 12 k$\lambda$.
(The $uv$ spacing in the whole data set ranges from 3.6 to 149.0 k$\lambda$.)

Figures 1(b) and 3 show the optimized map.
Some weak features in the FIR 5 and 6 areas became detectable.
Four compact sources were detected: FIR 5w, 5e, 6c, and 6n.
The 6.9 mm compact sources
are not associated with any IR/X-ray point source (Figure 1(b)).
Limiting the $uv$ space can cause some extended flux missing,
and excluding the data in the $uv$ space shorter than 12 k$\lambda$
would suppress structures larger than $\sim$17$''$.
Since the angular separation of each of the FIR 5/6 binary systems
is $\sim$4$''$ (much smaller than 17$''$),
it is unlikely that the measured flux densities of these compact sources
are significantly affected by the imaging procedure.

\begin{figure}[!t]
\epsscale{1.0}
\plotone{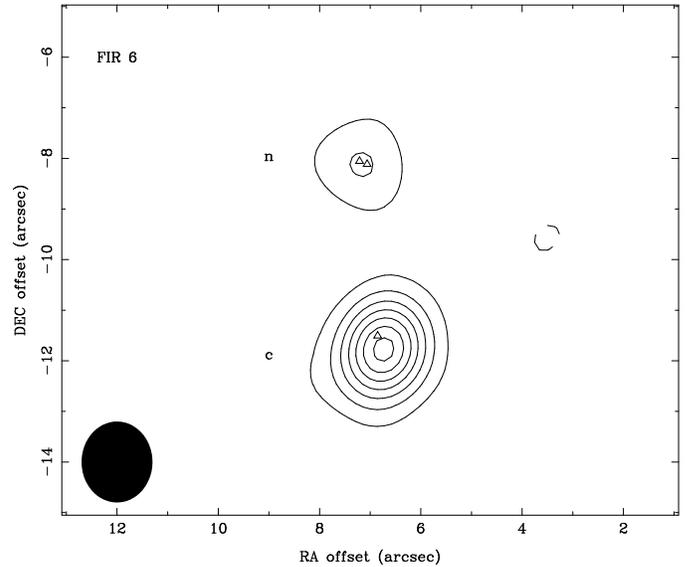}
\caption{\small\baselineskip=0.825\baselineskip
Map of the $\lambda$ = 6.9 mm continuum emission toward the FIR 6 region,
made with $R_{uv}^{\rm min}$ = 12 k$\lambda$ and a uniform weighting.
The contour levels are 1, 2, 3, 4, 5, 6, and 7 $\times$ 0.3 mJy beam$^{-1}$.
Shown in the bottom left-hand corner is the synthesized beam:
FWHM = 1\farcs6 $\times$ 1\farcs4 with P.A. = 0$^\circ$.
Open triangles:
H$_2$O maser sources.}
\end{figure}

\begin{deluxetable}{lccccc}
\tabletypesize{\small}
\tablecaption{NGC 2024 FIR 6 Water Maser Source Parameters}%
\tablewidth{0pt}
\tablehead{
&& \multicolumn{2}{c}{Peak Position\tablenotemark{a}}
& \colhead{$V_{\rm LSR}$} & \colhead{Peak Flux Density} \\
\cline{3-4}
\colhead{Source} & \colhead{Epoch}
& \colhead{$\alpha_{\rm J2000.0}$} & \colhead{$\delta_{\rm J2000.0}$}
& \colhead{(km s$^{-1}$)} & \colhead{(Jy beam$^{-1}$)}}%
\startdata
FIR 6c                  & 1996 Mar 15 & 05 41 45.147 & --01 56 04.01
                        & --8.4 & \phn0.32 $\pm$ 0.01  \\
FIR 6n\tablenotemark{b} & 1995 Nov 17 & 05 41 45.161 & --01 56 00.62
                        &  11.3 &    20.39 $\pm$ 0.02  \\
                        & 1996 Mar 15 & 05 41 45.171 & --01 56 00.56
                        &  11.3 &    67.21 $\pm$ 0.02  \\
\enddata\\
\tablenotetext{a}{Units of right ascension are hours, minutes, and seconds,
                  and units of declination are degrees, arcminutes,
                  and arcseconds.}%
\tablenotetext{b}{The peak velocity and flux densities
                  are from the smoothed spectra (Figure 4(b)--(c)).}%
\end{deluxetable}

To understand the effect of excluding the short-spacing data,
subregions of the map from the whole data set (Figure 1(a))
were fitted with two elliptical Gaussian objects:
a compact one representing a compact source
and an extended one representing the spurious stripe of the SCP.
The total flux of the compact object from the double Gaussian fit
would be larger than the corresponding value from the optimized map (Table 1),
if a significant amount of flux is missing
as a result of limiting the $uv$ space.
For the FIR 5 region, the double Gaussian fit
gives a total flux density of 4.8 mJy for the compact object (FIR 5w),
which is {\it smaller} than the value from the optimized map.
While the spurious negative stripe is affecting the flux measurement,
it is not clear if or not the measured flux of FIR 5w is reduced
as a result of the restriction of $uv$ space.
For the FIR 6 region, the double Gaussian fit
gives a total flux density of 4.1 mJy for the compact object (FIR 6c),
which agrees with the value from the optimized map within the uncertainty.
Therefore, we conclude that excluding the short-spacing data
did not significantly affect
the measured flux densities of the compact sources.

\subsection{Water Maser}

From the AC 443 data set,
H$_2$O maser sources were detected in the FIR 6 region.
Only one source was detected in 1995 November,
and it is associated with FIR 6n.
Two sources were detected in 1996 March,
one associated with FIR 6c and the other with FIR 6n.
The maser source parameters are listed in Table 2,
and the spectra are shown in Figure 4.

\begin{figure}[!t]
\epsscale{1.0}
\plotone{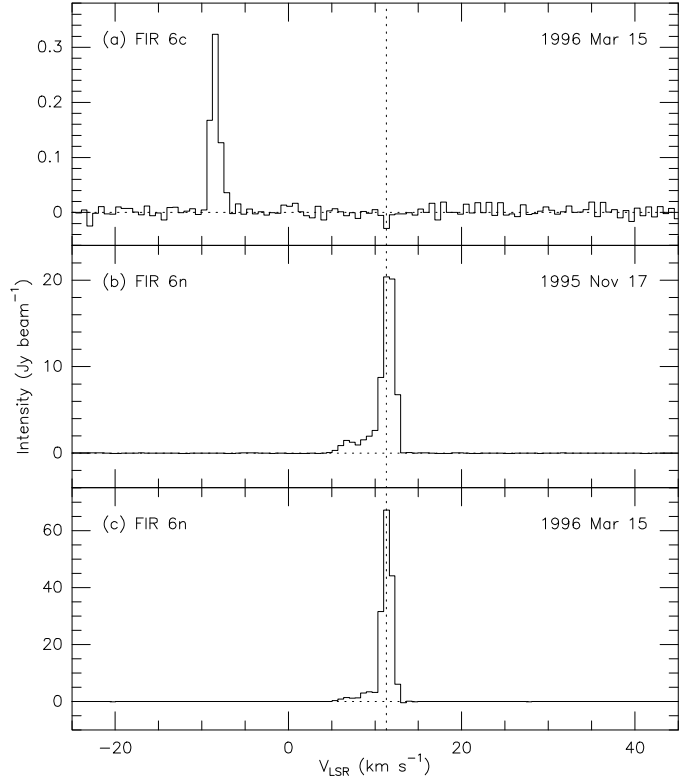}
\caption{\small\baselineskip=0.825\baselineskip
Spectra of the H$_2$O maser line.
(a)
Spectrum of FIR 6c from the C-array observations in 1996 March.
(b)
Spectrum of FIR 6n from the B-array observations in 1995 November.
(c)
Spectrum of FIR 6n from the C-array observations in 1996 March.
To reduce the ringing artifact,
the spectra of FIR 6n were smoothed
using a Hanning window over three channels.
Vertical dotted line:
velocity of the ambient dense gas
($V_{\rm LSR}$ = 11.3 km s$^{-1}$; Schulz et al. 1991).}
\end{figure}

The FIR 6c maser is relatively weak.
The maser peak position coincides
with the 6.9 mm continuum source position within 0\farcs4 (Figure 3).
This maser source has not been reported before.

The FIR 6n maser is much stronger than the FIR 6c maser.
The peak velocity of the FIR 6n maser in the two observing epochs
remained the same.
In addition to the narrow line core,
the spectra show a blueshifted line wing extending
down to $V_{\rm LSR} \approx$ 5 km s$^{-1}$ (Figure 4).

\section{DISCUSSION}

\subsection{FIR 6}

Both NGC 2024 FIR 6c and 6n appear compact in the 6.9 mm maps,
and they are well separated,
with little material surrounding them
(Figure 3; Lai et al. 2002; Alves et al. 2011).
The projected separation of FIR 6c--6n is 3\farcs9
or 1600 AU at a distance of 415 pc.
The FIR 6c--6n system was also marginally resolved
in the 450 $\mu$m continuum map of Visser et al. (1998),
and FIR 6c is stronger than 6n.

\subsubsection{FIR 6c}

Though FIR 6c is much brighter than 6n
in the millimeter and submillimeter range,
its nature is less clear.
There is no known outflow driven by FIR 6c.
Note that FIR 6c was not detected in the 3.6 cm continuum
while 6n was clearly detected (Rodr{\'\i}guez et al. 2003).
The detection of the H$_2$O maser at a velocity
blueshifted by $\sim$20 km s$^{-1}$ with respect to the ambient-gas velocity 
suggests that FIR 6c has an outflow activity.

The 6.9 mm continuum image (Figure 3) shows
that FIR 6c is marginally resolved.
A two-dimensional Gaussian fit gives a deconvolved source size
of FWHM = 0\farcs8 $\times$ 0\farcs2 with P.A. = --41$^\circ$,
which corresponds to 330 $\times$ 80 AU$^2$ at 415 pc.

\begin{figure*}[!t]
\epsscale{2}
\plotone{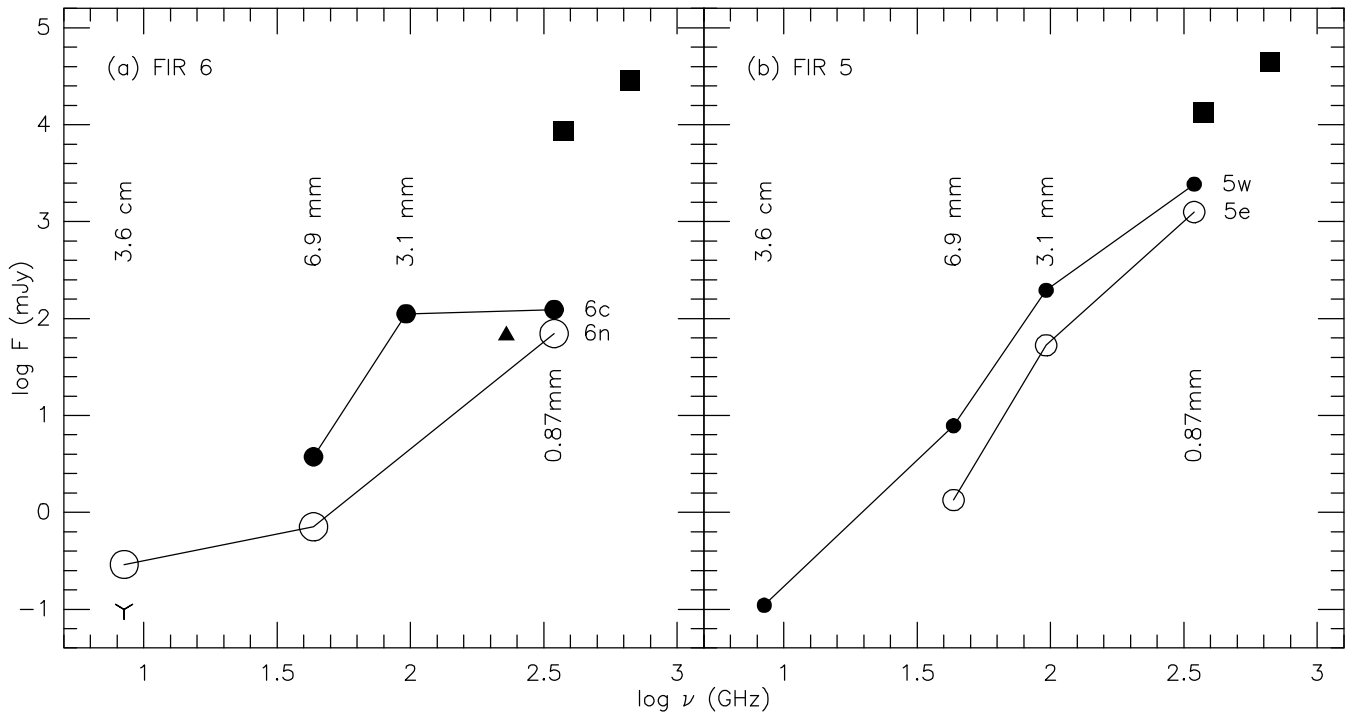}
\caption{\small\baselineskip=0.825\baselineskip
(a)
SEDs of FIR 6 sources.
Filled circles: FIR 6c.
Open circles: FIR 6n.
Filled squares: FIR 6 (including 6c and 6n).
Filled triangle:
peak flux density of FIR 6c at 1.3 mm (Lai et al. 2002).
Y sign:
upper limit for the flux density of FIR 6c at 3.6 cm.
(b)
SEDs of FIR 5 sources.
Filled circles: FIR 5w.
Open circles: FIR 5e.
Filled squares: FIR 5 (including 5w and 5e).
Flux densities are from Rodr{\'\i}guez et al. (2003), this work,
Wiesemeyer et al. (1997), Alves et al. (2011), and Visser et al. (1998).
The flux uncertainties are smaller than the size of markers.}
\end{figure*}

Figure 5(a) shows the spectral energy distribution (SED) of FIR 6c,
which is useful in understanding the origin of the 6.9 mm continuum emission.
SEDs can be described using a power-law form,
$F \propto \nu^\alpha$, where $F$ is the flux density,
$\nu$ is the frequency, and $\alpha$ is the spectral index.
The centimeter--millimeter SED of a YSO
usually has two components (Reynolds 1986; Anglada et al. 1998).
On the short wavelength side,
the emission comes from the dust in a disk/envelope system
and shows a large spectral index ($\alpha \gtrsim 2$).
On the long wavelength side,
the emission comes from ionized gas
(usually free--free emission from a thermal radio jet)
and shows a small spectral index ($\alpha \lesssim 1$).
At further longer wavelengths,
the free--free emission can become optically thick,
and the spectral slope can become steeper.

The slope in the (6.9 mm, 3.1 mm) section
is very steep: $\alpha \approx$ 4.3.
By contrast, the (3.1 mm, 0.87 mm) section is quite flat:
$\alpha \approx$ 0.1.
Such a flat spectrum around 1 mm is unusual for a YSO.
The 3.1 mm flux density of Wiesemeyer et al. (1997) cannot be very wrong
because it is consistent with those reported by Chandler \& Carlstrom (1996).
To check if the 0.87 mm flux of Alves et al. (2011) is in error,
the 1.3 mm flux of Lai et al. (2002) is plotted in Figure 5(a),
which confirms that the SED of FIR 6c is indeed flat.
Therefore, we conclude
that the millimeter SED of FIR 6c has a peculiar shape:
very steep on the long-wavelength side of the millimeter band
and very flat on the short-wavelength side.

It is unlikely
that continuum emission from dust
can display such a shallow slope around $\sim$1 mm.
Since the spectral index is not highly negative,
non-thermal radiation is also unlikely.
The most likely possibility is
that the millimeter continuum may come from free--free emission.
The break of slope near 3.1 mm
probably indicates a transition of optical depth.

Among the known types of YSOs,
only hypercompact (HC) H {\small II} regions can produce SEDs
similar to the one described above.
HC H {\small II} regions typically have a small size ($<$ 0.03 pc),
high electron density (10$^5$--10$^6$ cm$^{-3}$),
and high emission measure ($>$ 10$^8$ pc cm$^{-6}$;
Kurtz 2000; Beuther et al. 2007; Sewi{\l}o et al. 2011).
They are considered objects more evolved than a hot molecular core
and younger than an ultracompact H {\small II} region (Kurtz 2000).

Assuming that FIR 6c has a structure
similar to that of HC H {\small II} regions,
i.e., hot ionized gas confined by a dense molecular core,
interesting physical parameters can be derived.
For free--free emission,
the emission measure can be calculated by
\begin{equation}
   {\rm EM} \approx 12.1 \ \nu_0^{2.1} \ T_e^{1.35},
\end{equation}
where EM is in pc cm$^{-6}$, $\nu_0$ is the turnover frequency in GHz,
and $T_e$ is the electron temperature in K (Rohlfs \& Wilson 2006).
The SED of FIR 6c gives $\nu_0 \approx$ 100 GHz.
Assuming $T_e$ = 10$^4$ K,
the derived emission measure
is EM $\approx$ 5 $\times$ 10$^{10}$ pc cm$^{-6}$,
which is very high even for an HC H {\small II} region.
Given the 6.9 mm source size ($\sim$0\farcs4 or 7 $\times$ 10$^{-4}$ pc),
this EM implies an electron density of $\sim$8 $\times$ 10$^6$ cm$^{-3}$.

The high electron density implies
that the HC H {\small II} region should be
either confined by extremely dense molecular gas (with a density
about two orders of magnitude higher than the electron density)
or rapidly expanding.
The density of the molecular gas in the FIR 6 region
has so far been measured with single-dish observations.
Schulz et al. (1991) derived 4 $\times$ 10$^6$ cm$^{-3}$ for the CS-6 clump
using the CS $J$ = 5 $\rightarrow$ 4 and 7 $\rightarrow$ 6 lines.
Watanabe \& Mitchell (2008) derived 2 $\times$ 10$^6$ cm$^{-3}$
using the H$_2$CO $J_{K_{-1}K_{+1}}$ = 3$_{03}$ $\rightarrow$ 2$_{02}$
and 5$_{05}$ $\rightarrow$ 4$_{04}$ lines.
These estimates are lower limits of the density of the molecular gas
that may be immediately surrounding the FIR 6c HC H {\small II} region.
Interferometric observations in high-density tracer lines are needed
to directly investigate the presumed extremely dense molecular gas.
The detection of the H$_2$O maser suggests
that high-density ($\sim$10$^9$ cm$^{-3}$; Elitzur 1992) gas exists
in the FIR 6c region, at least locally.
The other possibility,
i.e., the rapid expansion of the HC H {\small II} region,
can be tested in the future
with high-resolution imaging observations over several years.

The number of ionizing photons can be calculated
using the equations given by Kurtz et al. (1994):
\begin{equation}
   N_c \approx 7.6 \times 10^{48} \ \nu^{0.1} \ T_e^{-0.5} \ S_\nu \ D^2,
\end{equation}
where $N_c$ is in photons s$^{-1}$, $\nu$ is the frequency in GHz,
$S_\nu$ is the flux density in Jy, and $D$ is the distance in kpc.
For FIR 6c, the flux is $\sim$0.12 Jy in the (3.1 mm, 0.87 mm) section,
and the derived Lyman continuum flux
is $N_c \approx$ 2.6 $\times$ 10$^{45}$ photons s$^{-1}$.
If this photon flux comes from a single zero-age-main-sequence star,
it would correspond to a B1 star
with a luminosity of $\sim$5800 $L_\odot$ and a mass of $\sim$13 $M_\odot$
(Panagia 1973; Lang 1992).
Such a high luminosity is not impossible,
considering that the total IR luminosity of the NGC 2024 complex
is 9.1 $\times$ 10$^4$ $L_\odot$ (Mezger et al. 1988).

It would be interesting to compare the mass of the central star
with that of the molecular clump/core containing FIR 6c.
However, the mass estimates in the literature have a large scatter,
depending on the tracers and assumptions
(see Section 1 of Watanabe \& Mitchell 2008).
Mezger et al. (1992) derived a mass of 8.3 $M_\odot$
for the molecular cloud core with a size of 12$''$ $\times$ 7$''$
based on the 870--1300 $\mu$m continuum.
Visser et al. (1998) derived 3.3 $M_\odot$
for the molecular core within a 14$''$ beam
based on the 800 $\mu$m continuum.
Watanabe \& Mitchell (2008) derived 2.0 $M_\odot$
for the core within a 15$''$ beam
based on the 850 $\mu$m continuum (Johnstone et al. 2006)
and the temperature derived from H$_2$CO lines.
Molecular line maps are more difficult to interpret
because the line peak positions are
sometimes different from the continuum peak position.
Schulz et al. (1991) derived a mass of $\sim$22 $M_\odot$
in a 20$''$ beam for the CS-6 molecular clump based on CS lines.
In summary, the estimated mass of the dense core
ranges from 2 to 22 $M_\odot$,
suggesting that calculations with simple assumptions
cannot easily capture the physical conditions of FIR 6c,
probably owing to the highly inhomogeneous environment.

Another peculiarity of FIR 6c is the extremely steep SED around 6.9 mm,
which is difficult to explain.
Other HC H {\small II} regions typically show
spectral slopes of $\alpha \approx$ 1 (Sewi{\l}o et al. 2011).
A possible answer would be a time variability of flux.
NGC 7538 IRS 1 is an example of HC H {\small II} region
displaying a time variability
(Franco-Hern{\'a}ndez \& Rodr{\'\i}guez 2004).
Sandell et al. (2009) suggested
that the H {\small II} region of IRS 1 is dominated by an ionized jet
and that the free--free emission from the jet is variable
because the accretion rate varies with time.
Future observations with higher angular resolutions
over a wide range of wavelength
would be helpful in understanding the nature of FIR 6c.
If FIR 6c is indeed an HC H {\small II} region containing a B1 star,
it would be one of the nearest examples of high-mass YSO
at this critical stage of evolution.

\subsubsection{FIR 6n}

Though the 6.9 mm total flux density of FIR 6n is smaller
than those of the other compact sources in this region,
FIR 6n seems to contain an active protostar.
FIR 6n drives a bipolar outflow in the east--west direction
(Richer 1990; Chandler \& Carlstrom 1996; Alves et al. 2011).
The 4.5 $\mu$m image indeed shows emission structures
on both east and west sides of FIR 6n (Figure 2).
The 4.5 $\mu$m emission may be tracing the outflow,
especially the outflow emitting in the H$_2$ line (Smith et al. 2006).

The H$_2$O maser position of FIR 6n
coincides with the 6.9 mm continuum position within $\sim$0\farcs1
(Figure 3; Genzel \& Downes 1977;
also listed as FIR 5 in Table 3 of Furuya et al. 2003).
The peak velocity of the maser in the 1995 and 1996 epochs
coincides with the ambient cloud velocity.
By contrast, the maser detected in 1999 February was at a different velocity,
redshifted by $\sim$5 km s$^{-1}$ (Furuya et al. 2003).
The H$_2$O maser is probably related with the outflow activity of FIR 6n.

The SED of FIR 6n (Figure 5(a)) is
shallow in the (3.6 cm, 6.9 mm) section ($\alpha \approx$ 0.6)
and steep in the (6.9 mm, 0.87 mm) section ($\alpha \approx$ 2.2).
This SED suggests
that the 3.6 cm continuum mostly comes from free electrons
and that the 0.87 mm continuum comes from dust.
The 6.9 mm flux is probably a combination of both.
The available data are too little to assess the dust properties.

\subsection{FIR 5}

The morphology of the dense gas in the NGC 2024 FIR 5 region
appears more complicated than that of FIR 6.
Lai et al. (2002) made a detailed image of the FIR 5 region
in the 1.3 mm continuum with a good sensitivity and $uv$ coverage.
They found that the FIR 5 core is separated into three components:
Main, NE, and SW (also see Alves et al. 2011).
The 6.9 mm image (Figure 1(b)) shows the FIR 5 Main component only.
Lai et al. (2002) resolved the Main component into seven clumps.
Only the two brightest ones, FIR 5w and 5e,
were detected in the 6.9 mm continuum.
The projected separation between them is 4\farcs9 (2000 AU).

\subsubsection{FIR 5w}

FIR 5w is the brightest millimeter continuum source
among the YSOs in the imaged region.
Several lines of evidence,
including the highly collimated outflow
(Richer et al. 1989, 1992; Alves et al. 2011),
suggest that FIR 5w contains an active protostar.

The 6.9 mm image (Figure 1(b)) shows
that FIR 5w is clearly an extended structure.
A two-dimensional Gaussian fit gives a deconvolved source size
of FWHM = 2\farcs2 $\times$ 1\farcs5 (910 $\times$ 620 AU$^2$)
with P.A. = 52$^\circ$.
The 6.9 mm source size is in good agreement
with that of the 0.87 mm source (Alves et al. 2011).

Figure 5(b) shows the SED of FIR 5w. 
The SED in the (3.6 cm, 0.87 mm) section
is steep ($\alpha \approx 2.7$) and steady,
suggesting that the 6.9 mm emission mostly comes from dust.
The SED is steep even in the (3.6 cm, 6.9 mm) section ($\alpha \approx 2.6$),
which suggests that a significant fraction of the 3.6 cm flux
may also come from dust.

With the SED of the dust component,
the mass of circumstellar molecular gas can be estimated
using Equation (1) of Chandler \& Carlstrom (1996)
and the mass emissivity given by Beckwith \& Sargent (1991).
In the (6.9 mm, 0.87 mm) section,
the SED gives an opacity index of $\beta \approx 0.6$.
Assuming a distance of 415 pc to the source and a dust temperature of 30 K
(Anthony-Twarog 1982; Chandler \& Carlstrom 1996),
the circumstellar gas mass of FIR 5w is $M$ = 0.6 $\pm$ 0.3 $M_\odot$,
which may include the inner protostellar envelope and the disk.
For comparison,
the mass of FIR 5w derived by Alves et al. (2011) is 1.09 $M_\odot$,
and the mass of the FIR 5 envelope within a 14$''$ beam
derived by Visser et al. (1998) is 5.1 $M_\odot$.

As discussed above, the 3.6 cm emission may not come from ionized gas,
which suggests
that the radio thermal jet of FIR 5w, if any, may be very weak.
This lack of evidence for radio jet is surprising
because FIR 5w drives a very well collimated molecular jet
(Richer et al. 1992; Alves et al. 2011).

\subsubsection{FIR 5e}

The nature of FIR 5e is less clear than that of FIR 5w.
There is so far no known outflow driven by FIR 5e.
Millimeter and submillimeter interferometric images
(Lai et al. 2002; Alves et al. 2011)
show that FIR 5e appears like a local peak
on a structure extended from FIR 5w.
The submillimeter polarimetry of Alves et al. (2011) showed
that the polarization direction of FIR 5e
is roughly perpendicular to that of FIR 5w,
which suggests that FIR 5e may be an object
with its own geometrical configuration,
separate from that of FIR 5w.
Detailed understanding of the FIR 5w--5e system
would require high-resolution images in the mid/far-IR bands,
but the bright nebulosity hampers the detection of these sources (Figure 2).

The steep SED of FIR 5e (Figure 5(b)) in the millimeter range
and the non-detection at 3.6 cm
suggest that most of the 6.9 mm emission comes from dust.
In the (6.9 mm, 0.87 mm) section,
the SED gives an opacity index of $\beta \approx 1.1$.
Using the procedure and parameters described in the previous section,
the mass of the dense molecular gas traced by the interferometric observations
is 0.5 $\pm$ 0.3 $M_\odot$.
For comparison,
the mass of FIR 5e derived by Alves et al. (2011) is 0.38 $M_\odot$.

\section{SUMMARY}

The NGC 2024 FIR 5/6 region was observed
using the VLA in the 6.9 mm continuum,
with an angular resolution of $\sim$1\farcs5,
to image the circumstellar structures of the YSOs.
In addition, an archival data set was analyzed
to investigate the H$_2$O emission sources associated with the YSOs.
The main results are summarized as follows.

1.
The 6.9 mm continuum maps show
compact emission sources associated with the YSOs
as well as an extended emission structure of the ionization front SCP.
To reduce the unwanted effect of the SCP on the neighboring areas,
the imaging process was optimized by limiting the $uv$ space.
In the optimized map, four compact sources were detected:
FIR 5w, 5e, 6c, and 6n.
The projected separations are
2000 AU for the FIR 5w--5e system and 1600 AU for the FIR 6c--6n system.
The source sizes are $\sim$760 AU for FIR 5w and $\sim$150 AU for FIR 6c.
The size of FIR 5e could not be determined well
because it is weak and located near FIR 5w.
FIR 6n is unresolved.

2.
The H$_2$O maser data show
that both FIR 6c and 6n exhibit maser activity.
The H$_2$O maser source associated with FIR 6c is relatively weak,
and the peak velocity is blueshifted by $\sim$20 km s$^{-1}$
with respect to the ambient cloud velocity.
The maser source associated with FIR 6n is strong,
and the peak velocity is very close to the ambient velocity.

3.
The SED of FIR 6c is steep around 6.9 mm and flat near 1 mm,
and the break of slope may be around 3 mm,
which is similar to the SED of HC H {\small II} regions.
If FIR 6c is an HC H {\small II} region,
i.e., a compact region of hot ionized gas heated by a massive YSO
and confined by a dense molecular cloud,
the emission measure is $\sim$5 $\times$ 10$^{10}$ pc cm$^{-6}$,
and the electron density is $\sim$8 $\times$ 10$^6$ cm$^{-3}$.
If this HC H {\small II} region is powered by a single object,
the central object may be a B1 star
with a luminosity of $\sim$5800 $L_\odot$ and a mass of $\sim$13 $M_\odot$.

4.
The SED of FIR 6n suggests
that the 6.9 mm continuum may be
a mixture of free--free emission and dust continuum emission.
The bipolar outflow and strong H$_2$O maser suggest
that FIR 6n contains an active protostar.

5.
The 6.9 mm continuum of FIR 5w seems to come from thermal dust emission.
The mass of the circumstellar gas is $\sim$0.6 $M_\odot$.
Though FIR 5w was suggested
to be the driving source of a strong molecular jet,
there is no evidence for detectable H$_2$O maser or radio thermal jet.

6.
The nature of FIR 5e is relatively less clear.
If it is a YSO, the mass of the circumstellar gas is $\sim$0.5 $M_\odot$.

\acknowledgements

M.C. and M.K. were supported by the Core Research Program
of the National Research Foundation of Korea (NRF)
funded by the Ministry of Education, Science and Technology (MEST)
of the Korean government (grant number 2011-0015816).
J.-E.L. was supported by the Basic Science Research Program
through NRF funded by MEST (grant number 2011-0004781).
The National Radio Astronomy Observatory is
a facility of the National Science Foundation
operated under cooperative agreement by Associated Universities, Inc.
This work is based in part on observations
made with the {\it Spitzer Space Telescope},
which is operated by the Jet Propulsion Laboratory,
California Institute of Technology, under a contract with NASA.
This publication makes use of data products
from the Two Micron All Sky Survey,
which is a joint project of the University of Massachusetts
and the Infrared Processing
and Analysis Center/California Institute of Technology,
funded by the National Aeronautics and Space Administration
and the National Science Foundation.


\begin{references}
\reference{} Alves, F. O., Girart, J. M., Lai, S.-P., Rao, R.,
             \& Zhang, Q. 2011, ApJ, 726, 63
\reference{} Anglada, G., Villuendas, E., Estalella, R., et al. 1998,
             AJ, 116, 2953
\reference{} Anthony-Twarog, B. J. 1982, AJ, 87, 1213
\reference{} Barnes, P. J., Crutcher, R. M., Bieging, J. H., Storey, J. W. V.,
             \& Willner, S. P. 1989, ApJ, 342, 883
\reference{} Beckwith, S. V. W., \& Sargent, A. I. 1991, ApJ, 381, 250
\reference{} Beuther, H., Leurini, S., Schilke, P., et al. 2007,
             A\&A, 466, 1065
\reference{} Chandler, C. J., \& Carlstrom, J. E. 1996, ApJ, 466, 338
\reference{} Chernin, L. M. 1996, ApJ, 460, 711
\reference{} Comer{\'o}n, F., Rieke, G. H., \& Rieke, M. J. 1996, ApJ, 473, 294
\reference{} Crutcher, R. M., Henkel, C., Wilson, T. L., Johnston, K. J.,
             \& Bieging, J. H. 1986, ApJ, 307, 302
\reference{} Elitzur, M. 1992, ARA\&A, 30, 75
\reference{} Franco-Hern{\'a}ndez, R., \& Rodr{\'\i}guez, L. F. 2004,
             ApJ, 604, L105
\reference{} Furuya, R. S., Kitamura, Y., Wootten, A., Claussen, M. J.,
             \& Kawabe, R. 2003, ApJS, 144, 71
\reference{} Gaume, R. A., Johnston, K. J., \& Wilson, T. L. 1992,
             ApJ, 388, 489
\reference{} Genzel, R., \& Downes, D. 1977, A\&AS, 30, 145
\reference{} Johnstone, D., Matthews, H., \& Mitchell, G. F. 2006,
             ApJ, 639, 259
\reference{} Kurtz, S., Churchwell, E., \& Wood, D. O. S. 1994, ApJS, 91, 659
\reference{} Kurtz, S. E. 2000, RevMexAA Ser. Conf., 9, 169
\reference{} Lai, S.-P., Crutcher, R. M., Girart, J. M., \& Rao, R. 2002,
             ApJ, 566, 925
\reference{} Lang, K. R. 1992, Astrophysical Data I. Planets and Stars
             (Berlin: Springer)
\reference{} Mezger, P. G., Chini, R., Kreysa, E., Wink, J. E.,
             \& Salter, C. J. 1988, A\&A, 191, 44
\reference{} Mezger, P. G., Sievers, A. W., Haslam, C. G. T., et al. 1992,
             A\&A, 256, 631
\reference{} Panagia, N. 1973, AJ, 78, 929
\reference{} Reynolds, S. P. 1986, ApJ, 304, 713
\reference{} Richer, J. S. 1990, MNRAS, 245, 24P
\reference{} Richer, J. S., Hills, R. E., \& Padman, R. 1992, MNRAS, 254, 525
\reference{} Richer, J. S., Hills, R. E., Padman, R.,
             \& Russell, A. P. G. 1989, MNRAS, 241, 231
\reference{} Rodr{\'\i}guez, L. F., G{\'o}mez, Y., \& Reipurth, B. 2003,
             ApJ, 598, 1100
\reference{} Rohlfs, K., \& Wilson, T. L. 2006, Tools of Radio Astronomy
             (Berlin: Springer)
\reference{} Sandell, G., Goss, W. M., Wright, M., \& Corder, S. 2009,
             ApJ, 699, L31
\reference{} Schulz, A., G{\"u}sten, R., Zylka, R., \& Serabyn, E. 1991,
             A\&A, 246, 570
\reference{} Sewi{\l}o, M., Churchwell, E., Kurtz, S., Goss, W. M.,
             \& Hofner, P. 2011, ApJS, 194, 44
\reference{} Skinner, S., Gagn{\'e}, M., \& Belzer, E. 2003, ApJ, 598, 375
\enlargethispage{-54\baselineskip}
\reference{} Smith, H. A., Hora, J. L., Marengo, M., \& Pipher, J. L. 2006,
             ApJ, 645, 1264
\reference{} Thronson, H. A., Jr., Lada, C. J., Schwartz, P. R., et al. 1984,
             ApJ, 280, 154
\reference{} Visser, A. E., Richer, J. S., Chandler, C. J., \& Padman, R. 1998,
             MNRAS, 301, 585
\reference{} Watanabe, T., \& Mitchell, G. F. 2008, AJ, 136, 1947
\reference{} Wiesemeyer, H., Guesten, R., Wink, J. E., \& Yorke, H. W. 1997,
             A\&A, 320, 287
\reference{} Wilson, T. L., Mehringer, D. M., \& Dickel, H. R. 1995,
             A\&A, 303, 840
\end{references}
\end{document}